\documentclass[%
 aip,
 amsmath,amssymb,
 reprint,%
]{revtex4-1}

\usepackage{graphicx}% Include figure files
\usepackage{dcolumn}% Align table columns on decimal point
\usepackage{bm}% bold math
\usepackage[utf8]{inputenc}
\usepackage[T1]{fontenc}
\usepackage{mathptmx}
\usepackage{booktabs}
\usepackage{lipsum}
\usepackage{xcolor}

\usepackage{amsmath}
\usepackage{comment}
\usepackage{soul}
\usepackage[labelformat=parens,labelsep=quad,skip=3pt]{caption}
\usepackage{graphicx}
\usepackage{textcomp}
\usepackage{hyperref}
\usepackage{url}

\newcommand{\rev}[1]{{\color{black}#1}}

\begin{document}

\preprint{AIP/123-QED}

\title{A New Remote Monitor and Control System Based on SigFox IoT Network}
%\title{A New System for Remote Monitor and Control of Experiments\\Based on the SigFox IoT Network}
% Force line breaks with \\

\author{Lorenzo Francesco Livi}
\affiliation{ 
Physics and Astronomy Dept., University of Florence, I-50019 Sesto Fiorentino, Italy
}%

\author{Jacopo Catani}%
 \email{jacopo.catani@ino.cnr.it.}
\affiliation{ 
National Institute of Optics - CNR (CNR-INO), I-50019 Sesto Fiorentino, Italy%\\This line break forced with \textbackslash\textbackslash
}%
 \affiliation{European Laboratory for NonLinear Spectroscopy (LENS), I-50019 Sesto Fiorentino, Italy}%Lines break automatically or can be forced with \\

\date{\today}% It is always \today, today,

\begin{abstract}
We describe a new, low-cost system designed to provide multi-sensor remote condition monitoring of modern scientific laboratories, as well as to allow users to perform actions from remote locations in case of detection of specified events. The system is battery operated and does not require the presence of a Local Area Network (LAN) or WiFi (which are typically not available in case of, e.g. power losses), as it exploits the growing infrastructure of Internet of Things (IoT) Low Power Wide Area Networks (LPWAN). In particular our system exploits the new SigFox ultra-narrow-bandwidth (UNB) infrastructure, and provides for a bidirectional link between the instrumentation and the remote user even in case of power line outages, which are among the most critical situations that a scientific laboratory can withstand. The system can detect the occurrence of predefined events in very short times, and either autonomously react with a series of predefined actions, \rev{also allowing a remote user to timely perform additional actions on the system} through an user-friendly smartphone application or via a browser interface. The system also embeds a novel power-loss detection architecture, which detects power line failures in less than 2 ms. We provide a full characterization of the prototype, including reaction times, connection latencies, sensors sensitivity, and power consumption. 
\end{abstract}

\maketitle

\begin{figure*}[htbp]
    \centering
    \includegraphics[width=1\textwidth]{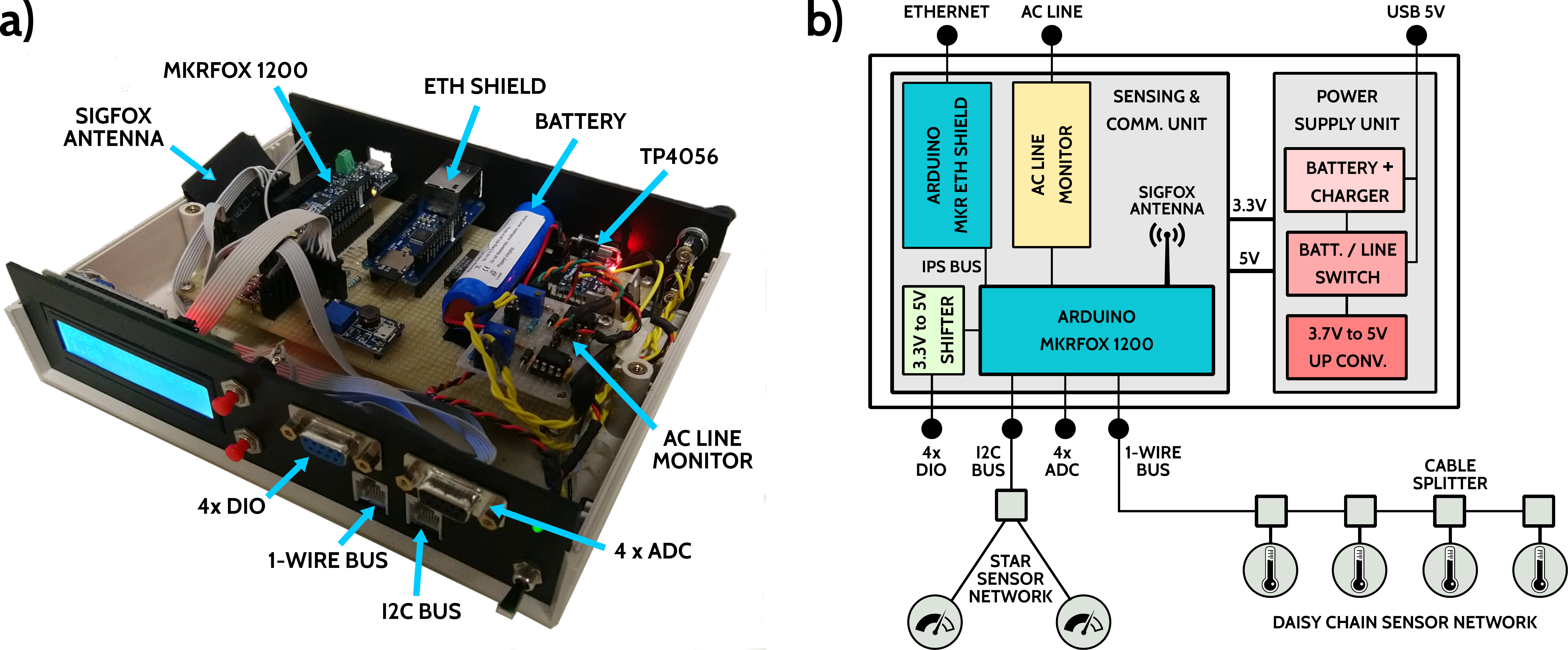}
        \vspace{2mm}
    \caption{\textbf{Prototypal board. a:} A photo of the final assembled prototypical board. \textbf{b:} The board conceptual schematics showing the main components of the power supply unit and of the sensing and communication unit. The sketch also shows a possible arrangement of the sensor network consisting of two sensors connected in "star" configuration to the I$^2$C bus and four thermometer probes connected in a "daisy chain" configuration to the 1-wire bus. See text for details.}
    \label{fig:board_photo_and_schematics}

\end{figure*}

\section{Introduction}\label{sec:intro}

Modern research laboratories, often involving delicate and sensitive instrumentation, are more and more highlighting the need for reliable remote condition monitoring and control capabilities \cite{labmonitoring2021}. Similar demands are also characterizing the design of modern industrial plants \cite{IndustrialMonitoring2006}, where the complexity of production processes poses the needs for optimized predictive maintenance routines to be achieved through the implementation of pervasive sensor networks \cite{IndustrialWirelessSensor2012}. Whilst in the industrial domain the main target of this deployment is the implementation of condition-based predictive maintenance routines with cost-effective sensors networks, in the case of scientific laboratories the aim of remote monitoring is to maintain the best experimental conditions to perform the most accurate measurement set, e.g., through stable environmental and/or instrumentation conditions\rev{. Another essential point concerns the prevention of} catastrophic failures of sensitive equipment (such as ovens, lasers, Ultra-high-vacuum (UHV) vessels and pumping stages) which can happen as a consequence of an unexpected critical event, such as, e.g, \rev{AC power line} blackouts and/or power, compressed air, vacuum line or technical gas supply failures.

Recently, development of complex remote sensing and monitoring systems has gained huge thrust from several sectors worldwide, due to the growing attention given to implementation pervasive sensing platforms, aimed at monitoring of a large number of endpoints with minimal cost \cite{pervasiveSensing2014}. In this scenario, Wireless Sensor Networks (WSNs) \rev{are} an emerging platform consisting of small-size, low-cost sensors with low energy requirements, that are used to sense, track measure, observe, and monitor mainly environmental phenomena, such as temperature, pressure, humidity, pollution, winds and send data wirelessly for processing \cite{WSNSurvey2014}. Among the 5G-compliant platforms which featured a rapid growth in the last years, the most prominent implementations of Low Power Wide Area Networks (LPWAN) are represented by Low-Power Long-Range Wide Area Networks (LoRaWAN),  Narrow-Band IoT (NB-IoT), Long-Term Evolution for Machines (LTE-M), and SigFox \cite{ComparativeStudyLPWAN2019, Survey5GEmergingTechnologies2019}. However, the possibilities offered by \rev{such} fast-growing low-power IoT network architectures \rev{could also be exploited in an extent wider than the pristine environmental monitoring}, to provide, e.g., a constant condition monitoring of scientific and industrial infrastructures not leveraging on local area networks, such as LAN and/or WiFi. This feature is essential in order to provide connectivity redundancy, as the latter are typically failing in case of power outages involving the building where laboratory and/or industrial equipment are located. This critical condition can put the apparatuses in an isolation condition, where no monitoring and/or control are possible even if local Uninterruptible Power Supplies (UPS) are present in lab, so that no proper countermeasures can be taken by remote users. Most of the LPWAN implementations mentioned above also offer a bidirectional link, which is particularly important as it allows remote users to have a certain degree of control on the monitored equipment. This is particularly relevant when human decisions could be required in addition to automated actions, as it could happen \rev{during the occurrence} of critical events. Moreover, in the particular case of scientific apparatuses or protected environments, where the electromagnetically-induced (EMI) noise can hamper the results of precision measurements, the possibility to exploit LPWANs \rev{avoids the need for} GSM/GPRS connection modules. \rev{This is an essential advance as these standards typically involve large EMI noise to be irradiated in the environment during transmission of data} \cite{CoverageComparisonGPRSvsLPWAN, CoverageAndCapacityAnalysis}. Furthermore, the latter are not meeting the low energy requirements of most recent IoT networks \cite{ComparativeStudyIOTvsCellular}, and battery operation can be an issue when long operation times are required.
Here, we implement and fully characterize a new multi-sensor platform for remote monitor and control of scientific, industrial apparatuses, based on the SigFox network (see Fig.\,\ref{fig:board_photo_and_schematics}), which has recently been deployed and represents one of the most promising LPWAN implementations for the development of pervasive sensing and actuation networks (see Sec.\,\ref{sec:sigFox}). \rev{Despite alternative IoT solutions currently available on the market (in particular LoRaWAN and NB-IoT) offer similar or even slightly better performance in terms of data rate and payload size, SigFox has been preferred over its direct competitors because of its excellent network coverage over the national territory and the availability of an Arduino board already equipped with a SigFox connectivity module, simplifying the implementation of the integrated remote sensing and control platform at both the software and hardware levels.} Differently from previous realizations \cite{WirelessSystemUltracoldAtoms}, our system provides a bidirectional wireless communication channel between remote users and the equipment under control, and features a highly versatile sensing architecture, based on low-cost, expandable set of I2C and 1-wire sensors, allowing for continuous monitoring of several parameters (temperature, pressure, humidity etc.). We also provide 4 general digital I/O ports, whose status can be changed by remote commands delivered through the SigFox network, allowing users to perform remote actions on the apparatus. The system also embeds a novel, low-cost AC line status and quality monitor, providing fast reaction capabilities (< 3 ms) to sudden variations of the AC line status. The system, which could run in battery mode for \rev{more than} 12 hours in our test campaign, performed inside a real scientific laboratory, could find applications in many sectors, ranging from remote condition monitoring of scientific research apparatuses and industrial plants, to on-field environmental and pollution monitoring.

\section{The SigFox network}
\label{sec:sigFox}
SigFox network \cite{SigFox_technology} is designed to provide a good coverage for either indoor and outdoor bi-directional long-range communications with low-powered, low-cost IoT devices. Long range and noise-resistant communication capability is achieved by means of a ultra-narrow band (UNB) transmission protocol which employs a strongly reduced transferring data rate (100 to 600 bps) to exchange 100 Hz wide messages over a narrow spectral sector (868 to 868.2 MHz in Europe) of the  publicly available band. Transmissions are always initiated by the remote device which remains in idle state for most of the time ensuring a high energy efficiency. 
Differently from other network architectures, this transmission scheme implies that communications between the devices and the base stations rely on a asynchronous exchange of messages. In particular, devices are not connected to a specific base station and each broadcast message is replicated three times over three different random frequencies and received by all the nearby stations. Due to the long transmission times, the UNB protocol employed makes it prohibitive to transmit large amount of data with battery powered devices so that a maximum of 12 bytes (8 bytes) payload size is allowed for uplink (downlink) messages. In addition to this constraint, governmental restrictions to the usage of the publicly available band limit the maximum number of daily message to 140 per device. The base stations receiving the device messages represent the first layer of the SigFox network architecture. The second other layer is constituted by the SigFox Support system, \rev{which is in charge of} processing of the received messages, their storage and final deployment. SigFox users can access message data exploiting a web interface or setting a callback to redirect the payload to a custom service and push an eventual downlink message to the IoT device.  \rev{SigFox network is administered by a single national operator,  providing a clear map of the network coverage, and does not require end-users to install and configure gateway units}.
Noticeably, the SigFox platform also features several collaborations with private partners (e.g. Arduino and Google) \cite{Arduino_MKRFOX1200_url, sigfoxgooglecollaboration}, and is expected to play a major role in the large-scale deployment of  pervasive WSNs and IoT services in the next few years. 

\section{System description}

The system architecture is based on a Arduino MKR FOX 1200 microcontroller \rev{board}, which is natively equipped with a SigFox connectivity module for bidirectional communications. The device connectivity also includes the standard set of I/O interfaces typical of the Arduino MKR series that comprises 7 analog inputs, 8 digital I/O, 1 DAC, a I$^2$C and a SPI communication bus \cite{Arduino_MKRFOX1200_url}. SigFox operations are enabled by an antenna tuned on the European ISM (Industrial, Scientific and Medical) frequency of 868 MHz and \textcolor{black}{accommodated inside the plastic box enclosing the board. Communications with the SigFox backend are characterized by an average Signal-to-Noise Ratio (SNR) between 10 and 24 dB and a RSSI (Received Signal Strength Indicator) between -101 and -131 dBm. 
%This choice grants the possibility to use oru system also 
Those values are measured with the board operating in indoor environments, where walls and concrete structures typically lower the RSSI by orders of magnitude with respect to outdoor configurations.} 
%\textcolor{red}{Siamo sicuri siano ordini di grandezza? Forse l'RSSI ma non il SNR. Vedi https://support.sigfox.com/docs/link-quality:-general-knowledge. Sigfox suggerisce di usare l'RSSI e non il SNR che non è un indicatore oggettivo.} Vedi link \hl{boooh vedi te come mettere Lorenzo}

As shown in Fig.\,\ref{fig:board_photo_and_schematics}b, where a modular scheme of the platform is provided, in addition to the MKR FOX 1200 microcontroller the board also includes an Arduino MKR ETH shield equipped with a Wiznet w5500 Ethernet controller chip. Due to the small payload size and limited number of messages daily allowed by the SigFox network, the system also embeds the possibility to exploit a wired Local Area Network (LAN) connection in order to share real time sensor data with remote users, leaving the SigFox connectivity available only for emergency or on-field conditions. Along with the w5500, the shield also hosts an SD card module that we employed to store sensor information and pre-loaded configuration parameters. \textcolor{black}{Both the devices, the w5500 and the SD module are interfaced to the Arduino exploiting the same \rev{SPI} bus.}
%As both the w5500 and the SD are interfaced exploiting the same Arduino IPS bus, two chip select pins are necessary to choose the device to communicate with.

In order to support versatile remote monitor applications, the board provides two general purpose I$^2$C and 1-Wire buses which allow for several low-cost and low-power typologies of \textcolor{black}{digital sensors for which Arduino libraries already exist} to be connected simultaneously in "star" or "daisy chain" arrangement with a minimal effort for the end-users. In addition to the extended assortment of commercial sensors available, the strength of the two buses consists in the minimal amount of wires needed to  control the connected devices\rev{. Indeed, only two wires and one wire (excluding the power connections) are employed by the I$^2$C and (as the name suggests) 1-Wire buses, respectively.} Being at most necessary four wires (two for communication, one for ground reference and one for the power supply) to control the sensors, we decided to use 4-poles telephonic cables and RJ11 6P4C connectors to assembly the sensor network, as this standard allows to build complex arrangements exploiting inexpensive components and cable splitters.
I$^2$C bus is directly connected to the dedicated pins on the Arduino microcontroller and supports 3.3 V sensors like the 16-bit ADC ADS1115 that we successfully tested. 1-Wire bus, instead, has been designed to work in normal power mode with a 3.3V dedicated supply line as this modality assures a better stability. Among the thermometry-related probes exploiting the 1-Wire technology, we successfully tested the Maxim DS18B20 temperature sensor and the Adafruit MAX31855 K-type thermocouple amplifier. 
%In order to facilitate the assembly of the sensor network, both the buses employ the RJ11 interface standard which allows to build complex arrangements exploiting inexpensive cable splitters. 

In order to expand the monitor and control capabilities of our system, the set of I/O connections available on the board also includes four 12-bit ADC and four general purpose digital I/O, both directly provided by the Arduino MKR FOX 1200 board. The digital I/Os could be either set as input or output terminals. As explained later in the text, the latter case is particularly relevant as it provides a remote user with the possibility to control an experimental apparatus or general equipment, through either the SigFox or LAN network. As the Arduino board uses a 3.3V supply, the digital I/O ports are provided with a 3.3V to 5V bidirectional level shifter 
%\hl{mettere codice} \textcolor{red}{Non c'è il codice. E' un circuitino preso su amazon con 4 mosfet e 8 resistenze, al limite si mette il modello di mosfet...}
in order to be compliant with the 5V TTL logic, widely used in both industry and laboratory settings.%Those digital I/O can be employed either to monitor a digital value or as an output for the execution of a remote command, as explained later in the text. 

The board is normally powered via an Universal Serial Bus (USB) connector, while a battery power supply has been included in order to ensure monitoring and connectivity functions even in case of blackout/failure of mains AC power line. The battery module features an inexpensive TP4056 USB single-cell Li-Ion charger and a MOSFET-based circuitry to automatically switch between USB and battery supply in the case of a power outage. Finally, a 3.3V low-dropout (LDO) regulator is used to deliver a constant voltage to the Arduino board and all the other subsystems, with the only exception being the SD-Ethernet shield and a I$^2$C system status display for which 5V supply is provided by means of a XL6009 DC-DC step-up converter. The power supply unit also provides the supply for the external sensors connected to the board. As shown in Table \ref{table:power_consumption}, the board power consumption (excluding the external sensors) strongly depends on the functioning condition, spanning from 125 mA  during normal monitoring operation, and reaching a maximum of 150 mA during SigFox transmission. Power saving strategies are adopted in emergency condition, when, e.g, 56 mA can be saved by commanding the Wiznet Ethernet chip to enter in power down mode in the case the board is running on battery and the internet connection is unavailable. 
\textcolor{black}{As energy consumption is one of the most challenging issues in WSN systems, additional power saving actions, e.g. the deactivation of the board display, makes it possible to sustain battery-powered board activities for several hours without recharging, allowing for extended outdoor operations. 
As an example, up to one day of full operation in emergency/outdoor conditions  is attainable before complete discharge with the 2600 mAh single-cell Li-Po battery that we embed in the board. }

\begin{table}[]
\centering

\begin{tabular}[t]{lc}
\toprule
&Current consumption\\
\midrule
Normal operating conditions & 125 mA \\
Arduino MKRFOX 1200 & 32 mA\\
Ethernet shield & 56 mA\\
SigFox downlink & 20 mA\\
SigFox uplink & 30 mA\\
\bottomrule
\end{tabular}
\caption{ Board components current consumption. The normal operating conditions refer to the board battery powered, the Ethernet shield on and no sensor connected. Additional power is needed for SigFox communication during uplink and downlink, respectively.}
\label{table:power_consumption}
\end{table}

\begin{figure*}[ht]
    \centering
    \includegraphics[width=1\textwidth]{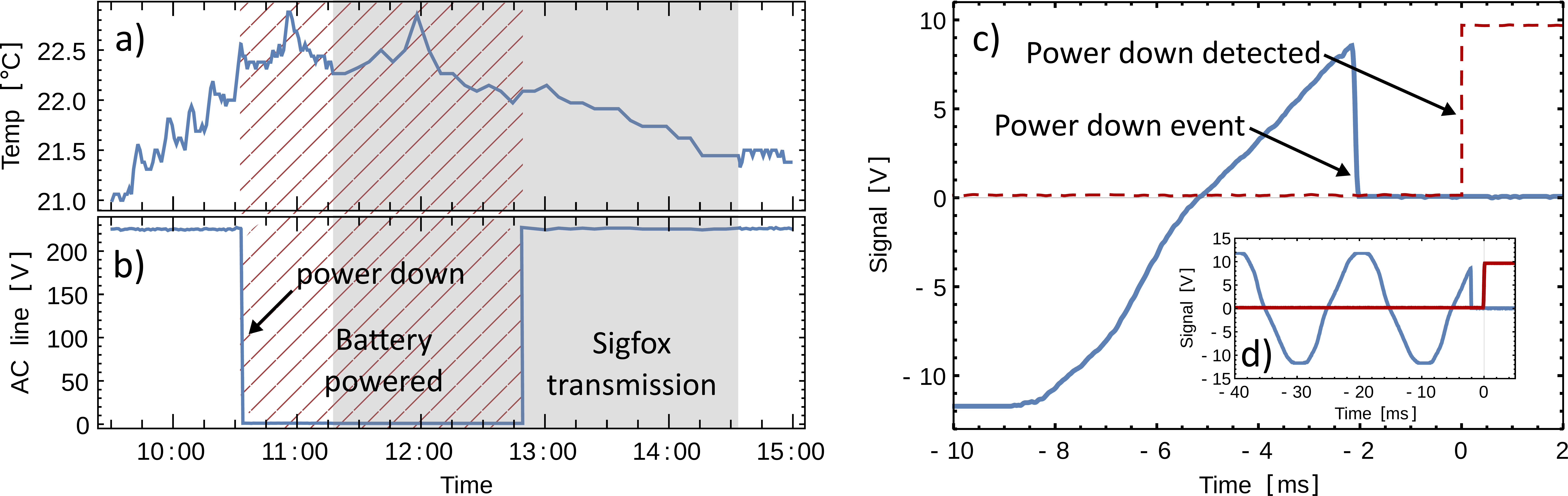}
    \caption{\textbf{Sensor data acquisition. a - b:} Example of sensor data log during a simulated power outage event. Panel a reports the log of a 1-Wire DS18B20 thermometer probe monitoring the laboratory ambient temperature while panel b reports the log of the main AC power line RMS value. The dashed and gray shadowed areas mark the time intervals during which the board was battery powered, and those in which SigFox communication was employed, respectively.
    At approximately 10:30 the board automatically switches to battery supply in response to a power outage. Later (approx. 11:15), a failure of the Ethernet connectivity forces the board to switch to SigFox network to communicate sensor data.  Sensor data update rate to the database is (1 min)$^{-1}$ during normal operating conditions and (5 min)$^{-1}$ when SigFox communication is employed.
    \textbf{c - d:} Typical board reaction time to a main AC line power outage. The blue line reports the sinusoidal AC RMS value in entrance to the board while the discontinuity of the red dashed line marks the time at which the board detect the power outage. In this example the detection time is 2.15 ms.    }
    \label{fig:sensor_data_analysis}
\end{figure*}

\section{Sensor initialization and data acquisition}
\label{sec:sensors}

As discussed before, the board can handle a non-predetermined number of different sensors whose settings are defined via a remote application. Each sensor connected to the board is identified by a series of properties that define its ID, model, physical address, operating and alarm ranges and conversion factors. This information is stored in a dedicated online database table which is updated exploiting the Android application, without the need for the end-user to change the Arduino firmware in order to modify the sensor properties, or add/remove sensors. There is no theoretical limit to the number of sensors that can be connected to the board, with the only exceptions of the Arduino flash memory required to store their information, and of the length of the cables connecting the sensor network to the board. \textcolor{black}{For both the 1-wire and I${^2}$C buses, maximum cable length strongly depends on the sensor network arrangement, number of sensors and maximum distance of the sensors from the board and spans from tens of centimeters in the case of I${^2}$C bus, up to tens of meters in the case of 1-wire connections \cite{1-wire_length}}.

As the number and properties of sensors are not hard-coded in the Arduino script but determined by the end-user via the Android application and stored online, two different procedures are exploited to initialize the board depending on the board operating condition. In normal operating conditions sensors settings are downloaded from the database table immediately after the board has been powered on and a copy is stored on the board SD card. In emergency/outdoor conditions, where the Ethernet connection is not available, the stored copy is employed during board initialization, so that no download from the SigFox network is needed. 
 
Once sensors are properly initialized, their read values are cyclically acquired and stored for sharing. The time between two subsequent readings of the same sensor is mainly limited by the long conversion time typical of the 1-Wire temperature sensors. The reading time depends on the chosen resolution of the 1-Wire sensor, which could be digitally set. Reading sequence times span from 94 ms at 9 bit resolution, up to 750 ms at 12 bit. This is not a particular issue as long as the board is exploited to monitor environmental variables or signals that are supposed to change on a timescale longer than the conversion time, e.g., ambient temperatures, pressures, humidity, or CW laser diodes output powers, pressure values in a vacuum vessels, or the polarization values out of a thermally-stressed single mode optical fiber.  
As an example of such a monitoring activity, Fig.\,\ref{fig:sensor_data_analysis}a reports the log of a DS18B20 temperature sensor during a simulated blackout and subsequent failure of both Ethernet connection and mains power supply, showing the capability of the board to support communication in emergency conditions with both USB and battery power. Sensor data have been acquired  at full resolution (which gives us a quantization level of 0.0625 \textdegree C) and setting a waiting time of 1.5 s between two acquisitions.  \textcolor{black}{The simultaneous log of the AC power line RMS value measured by the board dedicated ADC is reported in figure \ref{fig:sensor_data_analysis}b, with the gray shadowed area evidencing the time interval during which battery powering has been employed in response to the blackout event.}

\section{AC power line monitor and fast power loss detection}\label{sec:AC}

\begin{figure}[b]
    \centering
    \includegraphics[width=1\linewidth]{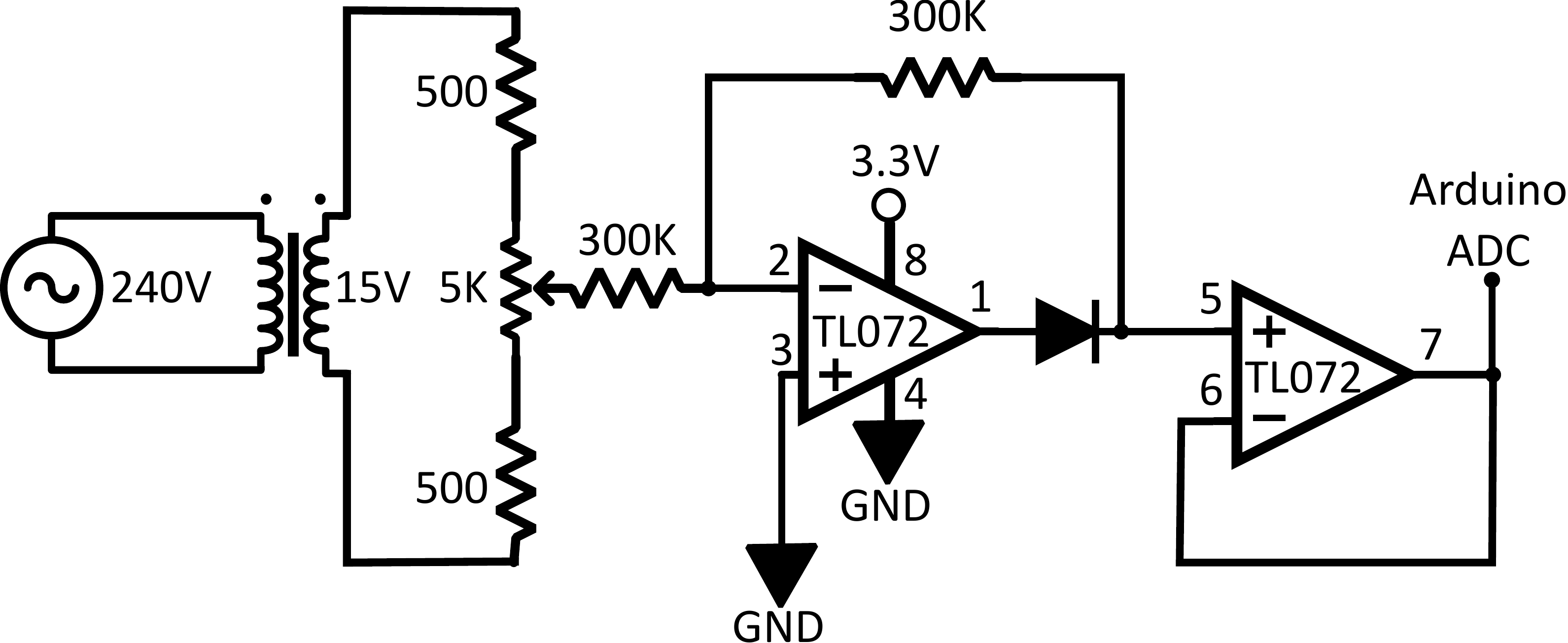}
    \caption{\textbf{AC power line monitoring schematics}: an AC-AC converter reduces the main electricity line voltage to a lower voltage which is fed into a precision rectifier in order to generate a positive-only periodic signal. This signal is sampled at regular times exploiting a dedicated ADC of the Arduino board in order to provide the current RMS voltage of the main line and detect eventual power outages.}
    \label{fig:AC_power_detector}
\end{figure}

Particular effort has been put to equip our system with a reliable sensor for AC power line monitoring and fast power outages / instabilities detection. AC power line monitoring constitutes a noticeable exception to the standard sensing and condition monitor scheme described above as, despite a few-ms power loss can be detrimental for laboratory instrumentation as well as for industrial equipment, this critical event would not be detected by our 1.5 s$^{-1}$ maximum sensor reading rate.  To this scope, as it is shown in Fig.\,\ref{fig:AC_power_detector}, the implemented sensing stage exploits an AC-AC converter to reduce the 240 VRMS of the main line to a safe lower voltage, and a precision rectifier to convert the sinusoidal waveform to positive-only values. The positive-only, periodic 50-Hz signal generated by the precision rectifier, is then processed by a dedicated  ADC channel of the Arduino microcontroller. For this purpose, in order to boost the ADC reading time, the ADC prescaler setting has been changed to 64 allowing a reduction of the sampling time from 430 $\mu$s to 21 $\mu$s \cite{adc_boost}. Fast real-time monitoring of the AC signal is then achieved by means of a software-based interrupt service routine performing a periodic analysis of the ADC read value every 2 ms. This time interval represents the lower limit to reaction times of our board to AC mains failures. Our optimized routine raises a digital alert every time the rectified signal drops below (above) a certain \rev{software} threshold value for more than 2 readings in a group of 3 consecutive, meaning that a power loss (power restore) is detected. Our AC monitor also provides a constant detailed reconstruction of the AC power line waveform, along with its RMS value. \rev{These features could also be exploited to detect more complex power line degradation events (e.g., power line brown-outs), occurring on timescales larger than few ms, after proper modification of the programming code uploaded on the microcontroller board}. 
%Fig.\,\ref{fig:sensor_data_analysis}c reports the reconstructed AC line waveform during an induced blackout event (solid blue line), 

\textcolor{black}{As an example of power\rev{-}down detection capability, Fig.\,\ref{fig:sensor_data_analysis}c-d reports the AC line waveform during an induced blackout event (solid blue line) along with the digital alert signal generated by the board as a consequence of the detected power loss (dashed red line), both measured with an oscilloscope probe. The figures evidence how our}
%, along with the digital alert signal generated by the board as a consequence of the detected power loss (dashed red line). As it is shown Fig.\,\ref{fig:sensor_data_analysis}c-d, this 
simple but effective AC monitor scheme allows for a reliable and stable detection of power outages / power restoration with minimal delays on the order of 2.5 ms, enabling for prompt automated reaction in case interlock procedures are required by the presence of delicate instrumentation. Power loss alert messages are also delivered to remote users for eventual further countermeasures or reaction through remote control of the board, enabled by our bidirectional communications scheme (see below). Alert messages, such as AC mains failure, are always uploaded via the SigFox network, given the high probability of Ethernet network failure as a contextual consequence of power outage.

\begin{figure}[b]
    \centering
    \includegraphics[width=1\linewidth]{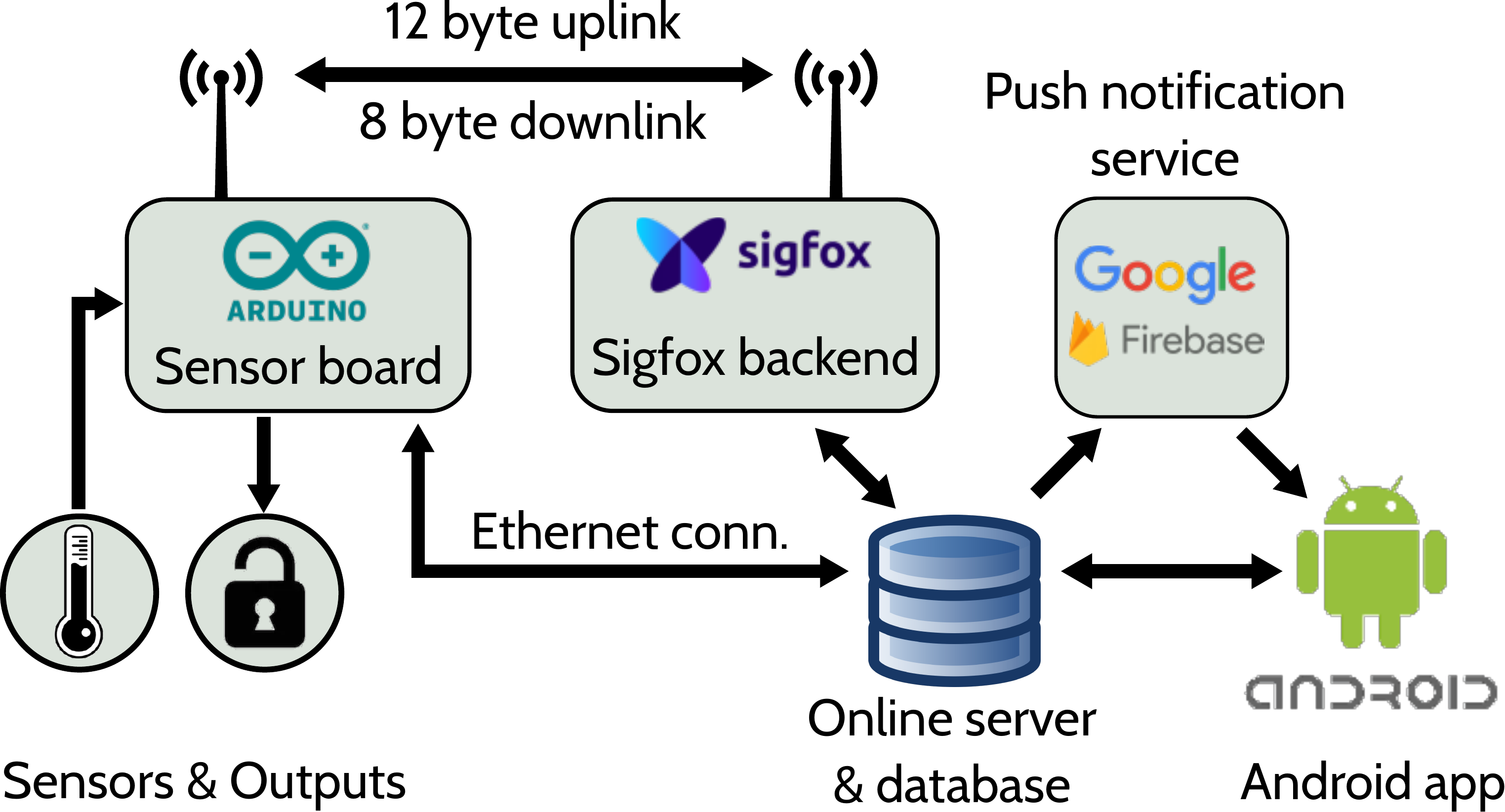}
    \caption{\textbf{Platform information flow}. Double head arrows indicate link bidirectional communication capability. Laboratory sensor data are transferred from the board to an online database via internet or SigFox connection, depending respectively whether the system is operating in normal or emergency conditions. Data can be accessed by the end-user via an Android application which also provides instantaneous visualization of critical events exploiting the Google Firebase push notification service. The links bidirectionality allows the user to set remote commands to be executed by the board. }
    \label{fig:information_flow}
\end{figure}

\vspace{-2mm}
\section{Information flow and communication latency}

As sketched in Fig.\,\ref{fig:information_flow}, the board is designed to work in two possible operating modes differing for the type of communication platform used to share the sensor data with the end-user and to remotely control the board, named \textit{normal} and \textit{emergency}, respectively. In normal operating conditions, data are acquired and transferred from the board to a web-based database via Ethernet connection. The database serves as an intermediate storage unit between the board and an Android application (see Sec.\,\ref{sec:android}) which provides end-users with services such as data visualization, remote control of the board through user-defined commands, or real-time definition of new sensors.  \rev{All the data transfer operations are managed by a PHP server hosted on a free web service which also hosts the MySQL database employed to store the sensor data and properties. } \textcolor{black}{An alternative data visualization modality exploiting a website is under development but not yet realized}.  Emergency conditions occur in the case Ethernet connectivity is not available and the board automatically switches to the SigFox network to transfer data to the database. Given the restrictions of the SigFox network in terms of transmission rate and payload size, different strategies have been adopted in order to share the sensor data in the two operating scenarios. During normal operation, sensor data sharing is realized exploiting a POST request to the database server through Ethernet. The refresh rate for data sharing can be set by user, but must not exceed the sensor data acquisition rate. When the Ethernet connectivity is lost (or in general with any user-defined event which can be configured by end-user) the board enters the emergency operation regime, and the sharing is performed via uplink through SigFox network. In this configuration, we have to cope up with the limit of 12 bytes in the payload size due to SigFox constrains, and with a maximum of 140 messages per day. This corresponds to a message every 11 minutes to evenly cover the whole day, but the messaging frequency could be increased, e.g., in case a more dense monitoring in correspondence of certain events or time periods is required. In order to maximize the information packed in a single SigFox message we engineered a packet frame structure which exploits the first byte of the payload to cast a code associated to a specific (user-defined) critical event, and arranges the values of 11 (user-defined) sensors in the remaining 11 bytes. As this data frame requires each sensor value to be converted to 8-bit resolution, we minimize information loss by shifting the conversion range (i.e. $2^8=256$ values) to a measurement range chosen by the user, which can be focused on the relevant variation range for each monitored quantity. Once the message is sent, a callback redirects the payload of uplink messages from the SigFox backend to the web server, which performs a conversion of the sensor values back to real units before storing them on the database. In both normal and emergency operation modes, end users can retrieve information on sensor readings through a dedicated Android application (see Sec.\,\ref{sec:android} for details).

\subsection{Remote control of the board through bidirectional data link}
\label{sec:bidirectional}
Besides representing a versatile and real-time remote condition monitoring system, our system offers a bidirectional data link in both Ethernet and SigFox connection configurations.  This feature allows our board to execute user-defined remote commands in either normal or emergency conditions, which has a prominent relevance in cases where human weighted decisions are preferable over automated / programmed countermeasures in response to critical events. This could encompass, e.g., fail-safe setting of equipment to preserve or restore its operating condition after power outages, or more generically any response to unexpected events for which no action could be defined before the critical event. As an example, a remote command, changing the status of one (or more) out the 4 digital I/O ports provided by our system can be employed to manually trigger an interlock, or to reactivate / reprogram an instrument after a blackout. 
When SigFox connectivity is employed a maximum of 4 downlink messages carrying a payload of 8 bytes can be received per day. Because of power saving requirements, inherited by the SigFox framework, downlink requests cannot be issued at every time, and they can only be performed after the device initiates the communication with an uplink message. For this reason the 8-bytes payload containing the remote commands list set by the user from the Android application is stored online and redirected to the device by the SigFox backend after the backend itself receives an uplink message. The downlink reception window starts 20 s after the uplink transmission has finished and lasts for 25 s. The whole operation consisting in an uplink followed by a downlink event takes about 40 s to complete, as shown in table \ref{table:system_reaction_times} where the reaction times of the system are reported. \textcolor{black}{In the normal configuration, being the board not continuously polling for downlink messages, a procedure similar to the one adopted in the emergency configuration is employed to retrieve user remote commands. In this configuration the amount of downlink messages is virtually unlimited, and command execution times are substantially limited by the latency of the Ethernet network+Arduino Ethernet shield, the command processing time of the software loaded on the Arduino board. The complete process requires typically less than 50 ms plus the Ethernet latency time.}

\begin{table}[]
\centering
\begin{tabular}[t]{lcc}
\toprule
& Absolute time & $\Delta$t \\
\midrule
Power down event & t$_0$ & -\\
Start alarm SigFox uplink & t$_0$ + 1.3 s & $\pm$0.5 s \\
Android notification displayed & t$_0$ + 4.5 s & $\pm$0.6 s \\
Remote SigFox command execution & t$_0$ + 39.6 s & $\pm$0.7 s \\
\bottomrule
\end{tabular}
\caption{System reaction times during emergency operating conditions. A power outage event at $t_0$ triggers the communication of an alarm exploiting the SigFox network. The board needs on average 1.3 s to compile the SigFox message and start the transmission. Alarm notification on the end-user device are displayed on average 4.5 s after the critical event. Finally, due to the long SigFox downlink times, the execution of a remote command takes about 40 s from the critical event to be started.}
\label{table:system_reaction_times}
\end{table}%

\section{Android application}
\label{sec:android}

\begin{figure}[t]
    \centering
    \includegraphics[width=1\linewidth]{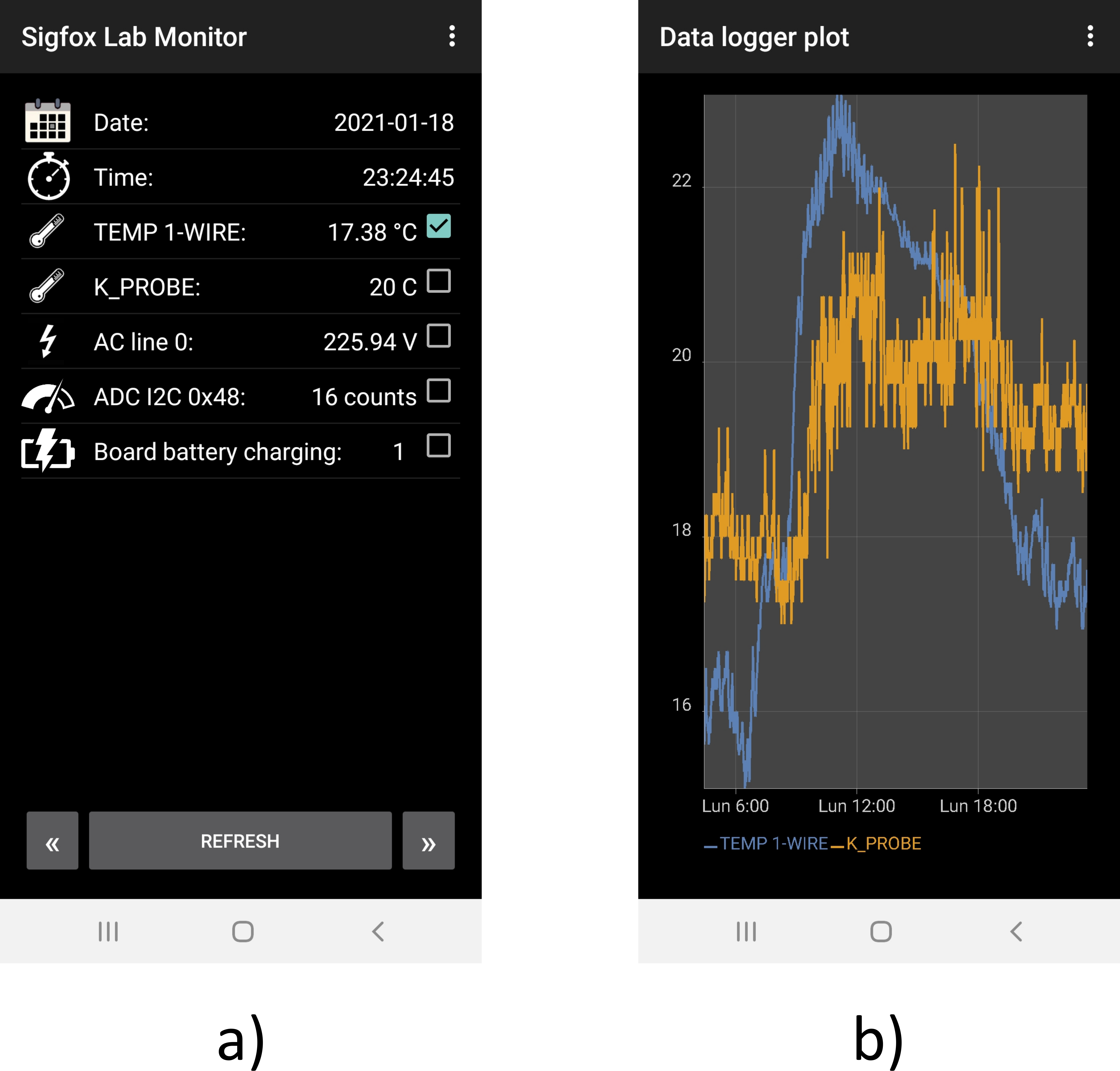}
    \caption{Examples of Android application screenshots showing the list of sensors connected to the board (\textbf{a}) and the time series plot of two temperature probes (\textbf{b}).}
    \label{fig:android_app}
\end{figure}

In the current version of the project, an Android application was chosen as the only interface between the end-user and the board. The extremely portable nature of this solution guarantees a constant monitoring of the sensor records as well as the possibility to react to unexpected critical event in any condition. The application has been developed within the Android Studio environment and registered on the Google Firebase platform in order to exploit its cloud messaging service \textcolor{black}{to notify user-defined, critical events such as a power outage or a sensor value exceeding a predefined threshold directly on the end-user device.} Fig.\,\ref{fig:android_app}a shows the main view of the application which reports a scrollable list of the board sensors together with their current value. Another view (see Fig.\,\ref{fig:android_app}b) gives the user the possibility to display a time series plot of one or multiple sensors \textcolor{black}{ as well as its time derivative with filtering options}. Additional views allows the user \textcolor{black}{to start monitoring new sensor attached to the board} or to edit the sensor properties, changing for example its name or the alarm ranges and to set a list of remote actions to be executed by the board. 
\textcolor{black}{Future application upgrades include the possibility to automatically scan the sensor network to find new sensors connected to the board, set time-programmed remote actions, control of multiple boards, export sensor data via social/email and the compatibility with IOS devices.}

\section{Conclusions}
In this work we presented a new system for remote \rev{monitoring} and control of experimental apparatuses based on the SigFox wireless communication network. Our system consists of a complete platform for multi-sensor data acquisition, bidirectional data sharing via either Ethernet WLAN or Sigfox network and data visualization on end-user mobile Android devices.
The data acquisition module is based on a battery-powered Arduino MKR FOX 1200 board  designed to sustain a complex, fully customizable general purpose sensor network operating exploiting the 1-wire and/or I$^2$C bus standards. Low power consumption and SigFox connectivity ensure a prolonged bidirectional link with remote instrumentation even in case of a power-outage of the electricity line or/and a failure of the WLAN network. An Android application provides the end-user with a portable interface to monitor sensor data and to plan the execution of remote commands. Alarm messages concerning out-of-threshold sensor values or emergency events like a blackout are notified directly on the user's device exploiting the Google Firebase cloud messaging service. The system has been tested with several low-cost commercially available sensors  and is equipped with a novel, custom AC power line monitor which demonstrated power outage detection times as low 2 ms. These features make our platform ideally suited for a plethora of applications which include continuous environmental and instrument variable monitoring, detection and real-time communication of critical events and remote manual intervention in case of instrument failure.

\begin{acknowledgments}
Authors would like to warmly thank L. Mischi for valuable technical support, and L. Fallani and G. Cappellini for carefully reading the manuscript. Authors would like to thank TechLab (http://quantumgases.lens.unifi.it/exp/tech) for financial and technological support to the project, and all members of Ytterbium team at LENS - Florence for valuable discussions and help during the test phase.
\end{acknowledgments}

\section*{Data Availability Statement}
Experimental data available on proper request from the authors. For further technical and availability information on the complete system please contact TechLab via e-mail: jacopo.catani@ino.cnr.it.

\bibliography{Bibliography}% Produces the bibliography via BibTeX.
%\externalbibliography{References}
%\bibliography{Bibliography}

\end{document}